\documentclass[aps,pra,showpacs,twocolumn,floatfix]{revtex4}
\usepackage{epsfig}
\usepackage{graphicx}
\usepackage{dcolumn}
\begin{document}
\title{Accurate determination of electric-dipole matrix elements in K and Rb
from Stark shift measurements}
\author{Bindiya Arora}
\affiliation{Department of Physics and Astronomy, University of Delaware, Newark, Delaware 19716}
\author{M. S. Safronova}
\affiliation{Department of Physics and Astronomy, University of Delaware, Newark, Delaware 19716}
\author{Charles W. Clark}
 \affiliation{
Physics Laboratory,
National Institute of Standards and Technology,
Technology Administration,
U.S. Department of Commerce, 
Gaithersburg, Maryland 20899-8410}
\date{\today} 
\begin{abstract}
Stark shifts of  potassium and rubidium  D1 lines have been 
measured with high precision by Miller {\it et al}
~\protect\cite{stk}. In this 
work, we 
combine these measurements with our all-order calculations to determine the values of the electric-dipole matrix elements for the 
$4p_{j}-3d_{j^{\prime}}$ transitions in K and
 for the $5p_{j}-4d_{j^{\prime}}$ transitions in Rb to high precision. The  $4p_{1/2}-3d_{3/2}$ and
 $5p_{1/2}-4d_{3/2}$ transitions contribute on the order of 90\% to the respective polarizabilities of the $np_{1/2}$ states in 
 K and Rb, and the remaining 10\% can be accurately calculated using the relativistic 
 all-order method.  Therefore, the combination of the experimental data and theoretical calculations
  allows us to determine the $np-(n-1)d$ matrix elements and their uncertainties.
  We compare these values with our all-order calculations of the  $np-(n-1)d$ matrix elements in K and Rb for a benchmark test of 
the accuracy of the all-order method for transitions involving $nd$ states.  Such matrix elements are of special interest for many 
applications, such  as determination of ``magic'' wavelengths in alkali-metal atoms for state-insensitive cooling and  trapping and 
determination of blackbody radiation shifts in optical frequency standards with ions.  
\end{abstract}
\maketitle

The values of the various electric-dipole matrix elements in alkali-metal atoms 
are needed for the variety of applications ranging from reducing decoherence in quantum 
logic gates 
\cite{Safronova:gate} to the study of fundamental symmetries 
 \cite{blundell:92,Vasilyev:beta}.
The all-order method that includes all single and double (SD) excitations of the Dirac-Fock (DF) 
wave function to all orders of perturbation theory was shown to give values for the primary $np-ns$ transitions in alkali-metal atoms
in excellent agreement with high-precision experiments 
\cite{relsd}. There are many
 interesting applications involving $np-n^{\prime}d$ transitions in alkali-metal atoms and other
 monovalent systems, but there are very few benchmark experiments, such as lifetime measurements, 
 to compare with theoretical calculations. Moreover, the only high-precision lifetime measurements  of the lowest 
 $nd$ states of any alkali-metal atom 
 \cite{5dT, 5dH}, carried out for Cs, are in disagreement 
 \cite{Safronova:cs} 
 with the  Stark shift values for cesium D1 and D2 lines 
 \cite{stk2,6pT}. 
  The breakdown of the correlation correction terms is very different 
 for the  $np-ns$ and $np-n^{\prime}d$ transitions, creating a need for additional benchmark tests. 
 In this work, we determine the values of the   $4p_{j}-3d_{j^{\prime}}$ transitions in K and
  $5p_{j}-4d_{j^{\prime}}$ transitions in Rb to high precision using experimental values of the 
  Stark shifts of D1 lines 
  \cite{stk} in these systems.
  
   The motivation for this work is two-fold. 
  First, we provide the recommended values for these transitions to be used for various applications, such
as determination 
of the ``magic'' wavelengths in alkali-metal atoms for 
state-insensitive cooling and  trapping and calculation of the $nd$ state polarizabilities. Second, we also
conduct all-order calculations of  these transitions in order to carry out a benchmark comparison of the
accuracy of the all-order method. The conclusions reached in this work 
allow us to provide recommended values for a variety of the transition properties of monovalent systems
and more accurately evaluate their uncertainties. 
Such transition properties are needed for  the evaluation of the black-body radiation and quadrupole shifts in ions, 
light-shifts and quadrupole polarizabilities in Ba$^+$ which were recently measured 
\cite{Fortson:ba,Lundeen:ba},
branching ratios for various decay channels, and other applications. Such benchmarks are also useful for the understanding of the accuracy of the all-order calculations conducted  for the  analysis of the experimental studies of parity 
violation with heavy atoms and search for an electron electric-dipole moment. 

 \begin{table*}
\caption{\label{tab1}  
The contributions to the
scalar polarizability for the $4p_{1/2}$ state in K and their uncertainties. The 
corresponding energy differences \cite{NIST1}
and the absolute values of the lowest-order (DF) and final all-order electric-dipole
 reduced matrix elements are also listed. The energy differences are given in cm$^{-1}$.
 Electric-dipole  matrix elements  are given in atomic units ($ea_0$), and 
 polarizabilities are given in $a_0^3$, where $a_0$ is the Bohr radius. }
\begin{ruledtabular}
\begin{tabular}{lrccrr}
\multicolumn{1}{c}{Contribution}&
\multicolumn{1}{c}{$k$}&
\multicolumn{1}{c}{$\langle 4p_{1/2} \|D\|k \rangle^{\rm{DF}}$}&
\multicolumn{1}{c}{$\langle 4p_{1/2} \|D\|k \rangle^{\rm{SD}}$}&
 \multicolumn{1}{c}{$E_{k}-E_{4p_{1/2}}$ }&
\multicolumn{1}{c}{$\alpha_0(4p_{1/2})$} \vspace{0.1cm}\\
\hline  \\[-0.4pc]
  $\alpha^{\text{main}}(ns)$& $4s$      &  4.555  &   4.102  &   -12985 &    -94.8(2)         \\
                            & $5s$      &  3.974  &   3.885  &     8041 &    137.3(1.4)       \\
                            & $6s$      &  0.925  &   0.903  &    14466 &      4.127(3)    \\
                            & $7s$      &  0.485  &   0.477  &    17289 &      0.962          \\
                            & $8s$      &  0.319  &   0.315  &    18780 &      0.386          \\
                            & $9s$      &  0.233  &   0.231  &    19663 &      0.198          \\
                            & $10s$     &  0.181  &   0.180  &    20229 &      0.117          \\
$\alpha^{\text{tail}}(ns)$         &           &        &         &        &    1.6(0.2)    \\[0.5pc]
  $\alpha^{\text{main}}(nd_{3/2})$  & $3d_{3/2}$& 8.596  &  7.949   & 8552  &  540.5(9.7)     \\
                                    & $4d_{3/2}$& 0.769  &  0.097   &14413  &    0.05(5)      \\
                                    & $5d_{3/2}$& 0.105  &  0.336   &17201  &    0.48(47)     \\
                                    & $6d_{3/2}$& 0.030  &  0.340   &18711  &    0.45(30)     \\
                                    & $7d_{3/2}$& 0.063  &  0.296   &19613  &    0.33(18)     \\
                                   &  $8d_{3/2}$& 0.069  &  0.253   &20193  &    0.23(11)     \\
                                   &  $9d_{3/2}$& 0.067  &  0.216   &20587  &    0.17(8)      \\
 $\alpha^{\text{tail}}(nd_{3/2})$  &            &  &         &        &    4.5(4.5)     \\[0.1pc]
$\alpha^{\text{core}}            $  &            &  &         &        &    5.5(3)     \\[0.1pc]
  Total                            &            &        &         &        & 602(11)   \\
\end{tabular}
\end{ruledtabular}
\end{table*}

The D1 line Stark shifts in alkali-metal atoms were measured with high precision  
by Miller
~\cite{stk}
 and Hunter {\it et al.}~
 \cite{stk1,stk2} using a pair of cavity-stabilized diode 
lasers locked to resonance signals. 
The K and Rb measurements, 39.400(5) kHz(kV/cm)$^{-2}$ and 61.153 kHz(kV/cm)$^{-2}$, respectively, represent a
three order of magnitude improvement in accuracy upon previous experimental results.
In this paper, we determine the 4$p_{1/2} - 3d_{3/2}$ and
$5p_{1/2} - 4d_{3/2}$ electric-dipole (E1) matrix elements in K 
and Rb, respectively, using these experimental Stark shifts. 
We also compare these values with our all-order 
calculations of the  $np-(n-1)d$ matrix elements 
in K and Rb for a benchmark test of the accuracy 
of the all-order method for transitions involving 
$nd$ states. 
The values of the $4p_{3/2} - 3d_{j}$ and
$5p_{3/2} - 4d_{j}$ electric dipole matrix elements in K and Rb, respectively, are obtained by 
combining our recommended values for the $4p_{1/2} - 3d_{3/2}$ and
$5p_{1/2} - 4d_{3/2}$ transitions with the corresponding accurate theoretical ratios. 

We start by expressing the experimental Stark shifts as the difference 
of the ground $ns$ and the first excited $np_{1/2}$ state polarizabilities of the respective atoms~\cite{stk}. 
It is convenient for this purpose to use the 
system of atomic units, a.u., in which $e, m_{\rm e}$, $4\pi 
\epsilon_0$ and the reduced Planck constant $\hbar$ have the 
numerical value 1.  Polarizability in a.u. has the dimensions of 
volume, and its numerical values presented here are thus measured 
in units of $a^3_0$, where $a_0\approx0.052918$~nm is the Bohr radius.
The atomic units for $\alpha$ can be converted to SI units via
 $\alpha/h$~[Hz/(V/m)$^2$]=2.48832$\times10^{-8}\alpha$~[a.u.], where
 the conversion coefficient is $4\pi \epsilon_0 a^3_0/h$ and the
 Planck constant $h$ is factored out. 
 
 \begin{table*}
\caption{\label{tab2}  The contributions to the
scalar polarizability for the $5p_{1/2}$ state in Rb and their uncertainties. The 
corresponding energy differences \cite{NIST}
and the absolute values of the lowest-order (DF) and final all-order electric-dipole
 reduced matrix elements are also listed. The energy differences are given in cm$^{-1}$.
 Electric-dipole  matrix elements  are given in atomic units ($ea_0$), and 
 polarizabilities are given in $a_0^3$, where $a_0$ is Bohr radius. }
\begin{ruledtabular}
\begin{tabular}{lrrrrr}
\multicolumn{1}{c}{Contribution}&
\multicolumn{1}{c}{$k$}&
\multicolumn{1}{c}{$\langle 5p_{1/2} \|D\|k \rangle^{\rm{DF}}$}&
\multicolumn{1}{c}{$\langle 5p_{1/2} \|D\|k \rangle^{\rm{SD}}$}&
 \multicolumn{1}{c}{$E_{k}-E_{5p_{1/2}}$ }&
\multicolumn{1}{c}{$\alpha_0(5p_{1/2})$} \vspace{0.1cm}\\
\hline  \\[-0.4pc]
  $\alpha^{\text{main}}(ns)$& $5s$      &  4.819    &   4.231   &  -12579   &  -104.11(15)          \\
                            & $6s$      &  4.256    &   4.146   &    7554   &   166.5(2.2)          \\
                            & $7s$      &  0.981    &   0.953   &   13733   &     4.835(16)      \\
                            & $8s$      &  0.514    &   0.502   &   16468   &     1.120(7)          \\
                            & $9s$      &  0.338    &   0.331   &   17920   &     0.448(3)          \\
                            & $10s$     &  0.247    &   0.243   &   18783   &     0.230(2)          \\
                            & $11s$     &  0.192    &   0.189   &   19338   &     0.135(1)          \\
$\alpha^{\text{tail}}(ns)$         &           &        &         &        &    1.9(0.2)    \\[0.5pc]
  $\alpha^{\text{main}}(nd_{3/2})$  & $4d_{3/2}$&  9.046 &  8.017    &   6777   &   694(30)       \\
                                    & $5d_{3/2}$&  0.244 &  1.352    &  13122   &  10.2(9)       \\
                                    & $6d_{3/2}$&  0.512 &  1.067    &  16108   &   5.2(1.1)       \\
                                    & $7d_{3/2}$&  0.447 &  0.787    &  17701   &   2.6(4)        \\
                                    & $8d_{3/2}$&  0.369 &  0.605    &  18643   &   1.4(2)        \\
                                   &  $9d_{3/2}$&  0.307 &  0.483    &  19243   &   0.89(10)       \\
 $\alpha^{\text{tail}}(nd_{3/2})$  &            &  &         &        &    10.5(10.5)     \\[0.1pc]
$\alpha^{\text{core}}            $  &            &  &         &        &    9.08(45)     \\[0.1pc]
  Total                            &            &        &         &        & 805(31)   \\
\end{tabular}
\end{ruledtabular}
\end{table*}
The Stark shifts in the D1 lines of K and Rb yield the following values of the
differences of the scalar dipole polarizability of the
$np_{1/2}$ and $ns$ states 
\cite{stk} :
	\begin{eqnarray} \label{eq-exp1}
	\Delta_{\rm{K}}  &=& \alpha_0(4p_{1/2})-\alpha_0(4s)=317.11(4)~a_0^3 \\
	\Delta_{\rm{Rb}} &=& \alpha_0(5p_{1/2})-\alpha_0(5s)=492.20(7)~a_0^3. \label{eq-exp2}
	\end{eqnarray} 
The static E1 atomic polarizabilities of the alkali-metal atoms are dominated by the valence contribution, 
$\alpha_v$, for all valence states. The small ionic core contribution evaluated in 
Ref.~\cite{RPA}
using random-phase approximation approach does not affect the present calculation as it is 
the same for $\alpha_0(np_{1/2})$ and $\alpha_0(ns)$ and does not contribute to the Stark shifts.
The counter term $\alpha_{vc}$ that needs to be introduced to correct ionic core polarizability 
 for an occupancy of the valence shell is very small for the $ns$ state and is entirely negligible for 
 the $np$ state. The valence polarizabilities are given in the sum-over-states approach by 
  \begin{equation}
\alpha_v = \frac{2}{3\mbox{(}2j_v+1\mbox{)}}\sum_k\frac{\left\langle k\left\|D\right\|v\right\rangle^2}{E_k-E_v}\label{eq-val},
\end{equation}
where $\langle k\|D\|v\rangle$ is the reduced electric-dipole matrix element for the 
transition between states $k$ and $v$, and $E_i$ is the energy corresponding to the level $i$.
The sum over the intermediate $k$ states converges very rapidly. 
In fact, the first two $ns-np_{1/2}$ and $ns-np_{3/2}$ transitions entirely dominate 
 the ground state polarizabilities. Since these values are known experimentally 
 \cite{Voltz},
our values for the ground state polarizabilities contain very little theoretical 
input. Such a calculation has been described before 
\cite{relsd, Derevianko:pol99,bin1,Safronova:8s07} 
and we do not repeat the details here. Our resulting values are $\alpha_0(4s)$=289.6(6)~$a_0^3$ and $\alpha_0(5s)$=318.35(62)~$a_0^3$. These results are in agreement with values of 
\cite{relsd, Derevianko:pol99}.
The uncertainty comes nearly entirely from the uncertainties of the experimental values of the 
$np$ lifetimes. 

The polarizabilities of the $np_{1/2}$ states are dominated by a single 
transition, $np_{1/2}-(n-1)d_{3/2}$, allowing us to use Eqs.~(\ref{eq-exp1}, \ref{eq-exp2})
to derive the matrix elements that are the subject of the present work. 
As a result, Eqs. (\ref{eq-exp1}, \ref{eq-exp2}) can be rewritten as 
	\begin{eqnarray} \label{eq1}
	\Delta_{\rm{K}}  &=& \alpha_0(4p_{1/2})-\alpha_0(4s) =
	\frac{1}{3} \frac{\left\langle 3d_{3/2}\left\|D\right\|4p_{1/2}\right\rangle^2}{E_{3d_{3/2}}-E_{4p_{1/2}}} \nonumber  \\
&	+&	\alpha^ {\rm{other}}_0(4p_{1/2}) -289.6(6) = 317.11(4)  \\
	\Delta_{\rm{Rb}} &=& \alpha_0(5p_{1/2})- \alpha_0(5s) =
		\frac{1}{3} \frac{\left\langle 4d_{3/2}\left\|D\right\|5p_{1/2}\right\rangle^2}{E_{4d_{3/2}}-E_{5p_{1/2}}} \nonumber  \\
&	+&	\alpha^{\rm{other}}_0(5p_{1/2}) - 318.35(62)  = 492.20(7),
	\end{eqnarray}
	where we substituted the ground state polarizability values and separated the contribution 
	from the $np_{1/2}-(n-1)d_{3/2}$ transition.  However, the remaining contributions to the 
	$np_{1/2}$ polarizabilities grouped
	together as  $\alpha^ {\rm{other}}_0(np_{1/2})$ still give 10 \% 
to the polarizabilities of the $np_{1/2}$ states and need to be evaluated accurately for our approach 
 to yield high-precision values. 	 
We describe such calculation below. For completeness, we describe the full theoretical evaluation of the 
$np_{1/2}$ polarizabilities first and then remove the dominate contribution to determine $\alpha^{\rm{other}}_0(np_{1/2})$ .
 
 We separate the valence polarizabilities into two parts, $\alpha^{\rm{main}}$, containing the 
 contributions from the states near the valence state, and the remainder $\alpha^{\rm{tail}}$. 
 We calculate the matrix elements contributing to the main term using the SD all-order  
method. We conduct additional semi-empirical scaling of our 
all-order values (SD$_{\rm{sc}}$)
where we expect scaled values to be more accurate 
based on the analysis of the dominant correlation correction contributions.
We refer the reader to 
Refs.~\cite{csao,blundell:92,relsd,usca} for the description of the 
all-order method and the scaling procedure. 
The experimental energies from \cite{NIST,NIST1} are used in all main term contributions. 
The remaining terms from highly-excited one-electron
states are included in the $\alpha^{\rm{tail}}$ part.
The $\alpha^{\rm{tail}}$ is calculated in DF approximation using a complete 
basis set functions that are linear combination of B-splines~\cite{johnson:Bspline88}.
We use 70 splines of order 11 for each angular momentum state. 
A spherical cavity radius of 220 a.u. is chosen to accommodate all 
valence orbitals included in the calculation of $\alpha^{\rm{main}}$.
We chose to include as many states as possible into $\alpha^{\rm{main}}$
in order to decrease the uncertainty in the remainder term. 

The contributions to the
scalar polarizabilities of the $4p_{1/2}$ state in K and $5p_{1/2}$ state
 in Rb and their uncertainties are listed in Tables~\ref{tab1} and \ref{tab2}, respectively. The 
corresponding experimental energy differences  
\cite{NIST-new,NIST,NIST1,rben}
and the absolute values of the lowest-order (DF) and final all-order electric-dipole
 reduced matrix elements are also listed. The lowest-order values are listed in order to illustrate
 the size of the correlation correction for all transitions. 
We use the experimental numbers from Ref.~\cite{Voltz} 
along with their uncertainties for the primary $ns-np_{1/2}$  transitions 
(for example, $5s-5p_{1/2}$  transition in Rb).
High-precision relativistic SD or scaled SD all-order values are used for all the remaining main term transitions. 
The uncertainties given for the matrix elements are 
equal to the differences between the SD \textit{ab initio} and scaled values.

The tail contribution is rather small for the $ns$ sum, but is significant for the $nd_{3/2}$ sum. 
In order to evaluate the uncertainties in the tail contributions, we calculated a few last main 
terms by using the DF approximation and compared the resulting values with the 
all-order values quoted in Tables \ref{tab1} and \ref{tab2}. In the case of the $ns$ states, the DF
values differ from our accurate values by only 7-10 \% with the difference decreasing with the 
principal quantum number $n$. As a result, we assign a 10 \% uncertainty to the $ns$ tail values. 
In the case of the $nd_{3/2}$ states, the correlation corrections are extremely  large and 
nearly cancel the lowest-order contribution for K. The differences between the all-order and DF
values are on the order of 90 \% for K and 60 \% for Rb. As we mentioned above, 
the large $nd_{3/2}$ tail contributions are the reason for the 
 inclusion of so many states into the main term. We take the uncertainty in the $nd_{3/2}$
 tail contributions to be 100 \%. 

 While such an estimate represents a high bound on the value of the Rb tail based on the 
 comparison of the DF and all-order values, the case of K requires some additional consideration
 owing to larger discrepancies of the DF and all-order values even for $n=9$.  
We have conducted additional all-order calculations of the K $4p_{1/2}-nd_{3/2}$  E1 matrix
elements with $n>9$. We found that our calculational scheme becomes impractical  for $n>15$.
Such a problem is expected because the lowest-order energies of the higher basis set states are large and 
positive resulting in cancellations in the denominators of the MBPT terms 
and consequently leading to the divergence of the 
all-order iteration procedure. In our approach, the tail does not sufficiently converge at $n=15$ to significantly reduce 
the tail error. As a result, we chose a different approach to ensure that we do not significantly 
underestimate the tail in the K calculation. We compared our final results for K and Rb with 
experimental values for the $np_{1/2}$ state polarizabilities  by combining the D1 line 
Stark shifts from \cite{stk} with recommended ground state polarizability values from  
\cite{Derevianko:pol99}. Our K and Rb values differ with the experimental results 606.7(6)~a.u. and 
810.6(6)~a.u. by 0.69 \% and 0.77 \%, respectively. If the tail in the K calculation were significantly 
underestimated, we should have seen a significant mismatch of the K and Rb comparison since the 
tail problem is specific to K calculation. 

  Tables ~\ref{tab1} and \ref{tab2} show that the uncertainties of the $nd_{3/2}$ tail 
 values give the overwhelmingly dominant contributions to the uncertainties of $\alpha^{\rm{other}}_0(np_{1/2})$.

  Subtracting the contribution from the $np_{1/2}-(n-1)d_{3/2}$
 states from our final theoretical values for $np_{1/2}$ polarizabilities and removing the corresponding uncertainties from the total error budget, we obtain
     	\begin{eqnarray*}
		\alpha^ {\rm{other}}_0(4p_{1/2}) & = & 61.6(4.8)~a^3_0  \\
	\alpha^ {\rm{other}}_0(5p_{1/2}) & = & 111(11)~a^3_0 . 
	\end{eqnarray*}
Substituting the $\alpha^ {\rm{other}}_0$ values and experimental energies from \cite{NIST-new} into
Eq.~(\ref{eq1}), we obtain the following absolute values of the E1 matrix elements: 
 \begin{eqnarray} \label{r1}
	\rm{K} && \langle 4p_{1/2} \|D\| 3d_{3/2} \rangle = 7.984(35)~ea_0 \\
	\rm{Rb} && \langle 5p_{1/2} \|D\| 4d_{3/2} \rangle = 8.051(63)~ea_0. \label{r2}
	\end{eqnarray}
	The uncertainties of these values are essentially defined by the uncertainties 
	of the respective $nd_{3/2}$ tail contributions for both K and Rb. The contributions of the uncertainties 
	in the Stark shift measurements and ground state polarizabilities to the 
uncertainty of the recommended matrix elements values given by (\ref{r1}, \ref{r2}) are negligible.
 
\begin{table}
\caption{\label{tab3} Comparison of the recommended values for the  $np_{1/2}-(n-1)d_{3/2}$ electric-dipole 
reduced matrix element in Rb and K 
derived from Stark shifts in this work (listed in row labeled ``Present work''), with our theoretical calculations carried out using
 single-double all-order method (SD), single-double 
all-order method including partial triple excitations (SDpT), and their scaled values. Absolute values are given. 
Units: $ea_0$. }
\begin{ruledtabular}
\begin{tabular}{lll}
\multicolumn{1}{c}{ } &
\multicolumn{1}{c}{K($4p_{1/2}-3d_{3/2}$)}&
\multicolumn{1}{c}{Rb($5p_{1/2}-4d_{3/2}$)}\\
\hline
 Present work               & 7.984(35)  & 8.051(63)  \\
 SD                                 & 7.868 & 7.846\\
 SD$_{\rm{sc}}$                     & 7.949 & 8.017\\ 
 SDpT                               & 7.956 & 7.994\\
 SDpT$_{\rm{sc}}$                   & 7.953 & 8.015\\ 
 Final theory                       & 7.949(80) & 8.02(17) \\
\end{tabular} 
\end{ruledtabular}
\end{table}

\begin{table} [t]
\caption{\label{tab4} Recommended absolute values of the $np-(n-1)d$ electric-dipole 
reduced matrix element in K and Rb. Units: $ea_0$. }
\begin{ruledtabular}
\begin{tabular}{llrll}
\multicolumn{2}{c}{K} &
\multicolumn{1}{c}{} &
\multicolumn{2}{c}{Rb}\\
\hline
   $4p_{1/2}-3d_{3/2}$& 7.984(35)&& $5p_{1/2}-4d_{3/2}$ &8.051(63) \\
   $4p_{3/2}-3d_{3/2}$& 3.580(16)&& $5p_{3/2}-4d_{3/2}$ &3.633(28) \\   
   $4p_{3/2}-3d_{5/2}$& 10.741(47)&& $5p_{3/2}-4d_{5/2}$ & 10.899(86)\\
\end{tabular} 
\end{ruledtabular}
\end{table}
We compare these final recommended results with our \textit{ab initio} and scaled  values in Table~\ref{tab3}. 
Since the contributions of the triple excitation are important for the accurate evaluation of these
matrix elements, we also conducted another all-order calculation partially including the triple 
excitations to the extent described in Ref.~\cite{relsd}. We refer to these results as SDpT values in  
Table~\ref{tab3} and text below. The corresponding scaled values are listed in row labeled ``SDpT$_{\rm{sc}}$''.
We take the SDpT$_{\rm{sc}}$ values as out final values (see, for example, Refs. \cite{usca,7d} for the discussion 
of this issue). We note that SD$_{\rm{sc}}$ and \textit{ab initio} SDpT values are essentially the same
in the case of K and very close in the case of Rb.  The uncertainty of the final values is taken to be the maximum difference between the final values and SD, SDpT, and SDpT$_{\rm{sc}}$ ones. Our all-order values are in excellent agreement with the values derived from the D1 line Stark shift. 
We also conclude that our procedure for the uncertainty evaluation of the theoretical matrix element 
values for the $np-(n-1)d$ transitions overestimates the uncertainty, 
especially for Rb. 

We also evaluated the recommended values of the $4p_{3/2}-3d_j$ E1 matrix elements in K and $5p_{3/2}-4d_j$ E1 matrix elements in Rb using our recommended values (\ref{r1}, \ref{r2}) and appropriate  
theoretical ratios. The ratios $\langle 4p_{1/2}\|D\|3d_{3/2}\rangle/ \langle 4p_{3/2}\|D\|3d_{3/2}\rangle$ and 
$\langle 4p_{3/2}\|D\|3d_{3/2}\rangle/\langle 4p_{3/2}\|D\|3d_{5/2}\rangle$ in K are essentially independent
of the correlation correction, i.e. the DF and all-order values are the same to better than 0.1\%. Therefore, 
the theoretical values of the ratio  are accurate enough for such a recalculation. The case of Rb is exactly the same as that of K. 
The complete set of our recommended values for all six $np-(n-1)d$ transitions considered in this work 
is given in Table~\ref{tab4}.  

In summary, relativistic all-order calculations of the $ns_{1/2}$ and $np_{1/2}$
state polarizability are presented. 
The calculations are combined with the experimental Stark shift
values in order to determine the $np_{1/2}-(n-1)d_{3/2}$ 
matrix elements in K and Rb atoms with high precision. 
The values of the matrix elements calculated using the experimental Stark shifts are found to be in
excellent agreement with the values of the reduced matrix elements 
evaluated using the relativistic all-order method. This work provides 
a benchmark test for the all-order matrix elements involving $nd$ states 
of monovalent systems. 

This research was performed
under the sponsorship of the National Institute of Standards
and Technology, US Department of Commerce.


\begin{thebibliography}{26}
\expandafter\ifx\csname natexlab\endcsname\relax\def\natexlab#1{#1}\fi
\expandafter\ifx\csname bibnamefont\endcsname\relax
  \def\bibnamefont#1{#1}\fi
\expandafter\ifx\csname bibfnamefont\endcsname\relax
  \def\bibfnamefont#1{#1}\fi
\expandafter\ifx\csname citenamefont\endcsname\relax
  \def\citenamefont#1{#1}\fi
\expandafter\ifx\csname url\endcsname\relax
  \def\url#1{\texttt{#1}}\fi
\expandafter\ifx\csname urlprefix\endcsname\relax\def\urlprefix{URL }\fi
\providecommand{\bibinfo}[2]{#2}
\providecommand{\eprint}[2][]{\url{#2}}

\bibitem[{\citenamefont{Miller et~al.}(1994)\citenamefont{Miller, Krause, and
  Hunter}}]{stk}
\bibinfo{author}{\bibfnamefont{K.~E.} \bibnamefont{Miller}},
  \bibinfo{author}{\bibfnamefont{D.}~\bibnamefont{Krause}}, \bibnamefont{and}
  \bibinfo{author}{\bibfnamefont{L.~R.} \bibnamefont{Hunter}},
  \bibinfo{journal}{Phys. Rev. A} \textbf{\bibinfo{volume}{49}},
  \bibinfo{pages}{5128} (\bibinfo{year}{1994}).

\bibitem[{\citenamefont{Safronova et~al.}(2003)\citenamefont{Safronova,
  Williams, and Clark}}]{Safronova:gate}
\bibinfo{author}{\bibfnamefont{M.~S.} \bibnamefont{Safronova}},
  \bibinfo{author}{\bibfnamefont{C.~J.} \bibnamefont{Williams}},
  \bibnamefont{and} \bibinfo{author}{\bibfnamefont{C.~W.} \bibnamefont{Clark}},
  \bibinfo{journal}{Phys.\ Rev.\ A} \textbf{\bibinfo{volume}{67}},
  \bibinfo{pages}{040303(R)} (\bibinfo{year}{2003}).

\bibitem[{\citenamefont{Blundell et~al.}(1992)\citenamefont{Blundell,
  Sapirstein, and Johnson}}]{blundell:92}
\bibinfo{author}{\bibfnamefont{S.~A.} \bibnamefont{Blundell}},
  \bibinfo{author}{\bibfnamefont{J.}~\bibnamefont{Sapirstein}},
  \bibnamefont{and} \bibinfo{author}{\bibfnamefont{W.~R.}
  \bibnamefont{Johnson}}, \bibinfo{journal}{Phys.\ Rev.\ D}
  \textbf{\bibinfo{volume}{45}}, \bibinfo{pages}{1602} (\bibinfo{year}{1992}).

\bibitem[{\citenamefont{Vasilyev et~al.}(2002)\citenamefont{Vasilyev, Savukov,
  Safronova, and Berry}}]{Vasilyev:beta}
\bibinfo{author}{\bibfnamefont{A.~A.} \bibnamefont{Vasilyev}},
  \bibinfo{author}{\bibfnamefont{I.~M.} \bibnamefont{Savukov}},
  \bibinfo{author}{\bibfnamefont{M.~S.} \bibnamefont{Safronova}},
  \bibnamefont{and} \bibinfo{author}{\bibfnamefont{H.~G.} \bibnamefont{Berry}},
  \bibinfo{journal}{Phys.\ Rev.\ A} \textbf{\bibinfo{volume}{66}},
  \bibinfo{pages}{020101(R)} (\bibinfo{year}{2002}).

\bibitem[{\citenamefont{Safronova et~al.}(1999)\citenamefont{Safronova,
  Johnson, and Derevianko}}]{relsd}
\bibinfo{author}{\bibfnamefont{M.~S.} \bibnamefont{Safronova}},
  \bibinfo{author}{\bibfnamefont{W.~R.} \bibnamefont{Johnson}},
  \bibnamefont{and}
  \bibinfo{author}{\bibfnamefont{A.}~\bibnamefont{Derevianko}},
  \bibinfo{journal}{Phys.\ Rev.\ A} \textbf{\bibinfo{volume}{60}},
  \bibinfo{pages}{4476} (\bibinfo{year}{1999}).

\bibitem[{\citenamefont{DiBerardino et~al.}(1998)\citenamefont{DiBerardino,
  Tanner, and Sieradzan}}]{5dT}
\bibinfo{author}{\bibfnamefont{D.}~\bibnamefont{DiBerardino}},
  \bibinfo{author}{\bibfnamefont{C.~E.} \bibnamefont{Tanner}},
  \bibnamefont{and}
  \bibinfo{author}{\bibfnamefont{A.}~\bibnamefont{Sieradzan}},
  \bibinfo{journal}{Phys.\ Rev.\ A} \textbf{\bibinfo{volume}{57}},
  \bibinfo{pages}{4204} (\bibinfo{year}{1998}).

\bibitem[{\citenamefont{Hoeling et~al.}(1996)\citenamefont{Hoeling, Yeh,
  Takekoshi, and Knize}}]{5dH}
\bibinfo{author}{\bibfnamefont{B.}~\bibnamefont{Hoeling}},
  \bibinfo{author}{\bibfnamefont{J.~R.} \bibnamefont{Yeh}},
  \bibinfo{author}{\bibfnamefont{T.}~\bibnamefont{Takekoshi}},
  \bibnamefont{and} \bibinfo{author}{\bibfnamefont{R.~J.} \bibnamefont{Knize}},
  \bibinfo{journal}{Opt.\ Lett.} \textbf{\bibinfo{volume}{21}},
  \bibinfo{pages}{74} (\bibinfo{year}{1996}).

\bibitem[{\citenamefont{Safronova and Clark}(2004)}]{Safronova:cs}
\bibinfo{author}{\bibfnamefont{M.~S.} \bibnamefont{Safronova}}
  \bibnamefont{and} \bibinfo{author}{\bibfnamefont{C.~W.} \bibnamefont{Clark}},
  \bibinfo{journal}{Phys.\ Rev.\ A} \textbf{\bibinfo{volume}{69}},
  \bibinfo{pages}{040501(R)} (\bibinfo{year}{2004}).

\bibitem[{\citenamefont{Hunter et~al.}(1992)\citenamefont{Hunter, Krause,
  Miller, Berkeland, and Boshier}}]{stk2}
\bibinfo{author}{\bibfnamefont{L.~R.} \bibnamefont{Hunter}},
  \bibinfo{author}{\bibfnamefont{D.}~\bibnamefont{Krause}},
  \bibinfo{author}{\bibfnamefont{K.~E.} \bibnamefont{Miller}},
  \bibinfo{author}{\bibfnamefont{D.~J.} \bibnamefont{Berkeland}},
  \bibnamefont{and} \bibinfo{author}{\bibfnamefont{M.~G.}
  \bibnamefont{Boshier}}, \bibinfo{journal}{Opt. Commun.}
  \textbf{\bibinfo{volume}{94}}, \bibinfo{pages}{210} (\bibinfo{year}{1992}).

\bibitem[{\citenamefont{Tanner and Wieman}(1988)}]{6pT}
\bibinfo{author}{\bibfnamefont{C.~E.} \bibnamefont{Tanner}} \bibnamefont{and}
  \bibinfo{author}{\bibfnamefont{C.}~\bibnamefont{Wieman}},
  \bibinfo{journal}{Phys.\ Rev.\ A} \textbf{\bibinfo{volume}{38}},
  \bibinfo{pages}{162} (\bibinfo{year}{1988}).

\bibitem[{\citenamefont{Sherman et~al.}(2005)\citenamefont{Sherman, Koerber,
  Markhotok, Nagourney, and Fortson}}]{Fortson:ba}
\bibinfo{author}{\bibfnamefont{J.~A.} \bibnamefont{Sherman}},
  \bibinfo{author}{\bibfnamefont{T.~W.} \bibnamefont{Koerber}},
  \bibinfo{author}{\bibfnamefont{A.}~\bibnamefont{Markhotok}},
  \bibinfo{author}{\bibfnamefont{W.}~\bibnamefont{Nagourney}},
  \bibnamefont{and} \bibinfo{author}{\bibfnamefont{E.~N.}
  \bibnamefont{Fortson}}, \bibinfo{journal}{Phys.\ Rev.\ Lett.}
  \textbf{\bibinfo{volume}{94}}, \bibinfo{pages}{243001}
  (\bibinfo{year}{2005}).

\bibitem[{\citenamefont{Snow et~al.}(2005)\citenamefont{Snow, Gearba, Komara,
  Lundeen, and Sturrus}}]{Lundeen:ba}
\bibinfo{author}{\bibfnamefont{E.~L.} \bibnamefont{Snow}},
  \bibinfo{author}{\bibfnamefont{M.~A.} \bibnamefont{Gearba}},
  \bibinfo{author}{\bibfnamefont{R.~A.} \bibnamefont{Komara}},
  \bibinfo{author}{\bibfnamefont{S.~R.} \bibnamefont{Lundeen}},
  \bibnamefont{and} \bibinfo{author}{\bibfnamefont{W.~G.}
  \bibnamefont{Sturrus}}, \bibinfo{journal}{Phys.\ Rev.\ A}
  \textbf{\bibinfo{volume}{71}}, \bibinfo{pages}{022510}
  (\bibinfo{year}{2005}).

\bibitem[{\citenamefont{Ralchenko et~al.}(2005)\citenamefont{Ralchenko, Jou,
  Kelleher, Kramida, Musgrove, Reader, Wiese, and Olsen}}]{NIST1}
\bibinfo{author}{\bibfnamefont{Y.}~\bibnamefont{Ralchenko}},
  \bibinfo{author}{\bibfnamefont{F.~C.} \bibnamefont{Jou}},
  \bibinfo{author}{\bibfnamefont{D.~E.~.} \bibnamefont{Kelleher}},
  \bibinfo{author}{\bibfnamefont{A.~E.} \bibnamefont{Kramida}},
  \bibinfo{author}{\bibfnamefont{A.}~\bibnamefont{Musgrove}},
  \bibinfo{author}{\bibfnamefont{J.}~\bibnamefont{Reader}},
  \bibinfo{author}{\bibfnamefont{W.~L.} \bibnamefont{Wiese}}, \bibnamefont{and}
  \bibinfo{author}{\bibfnamefont{K.}~\bibnamefont{Olsen}},
  \emph{\bibinfo{title}{Nist atomic spectra database}} (\bibinfo{year}{2005}),
  \bibinfo{note}{(version 3.1.2). [Online]. Available:
  http://physics.nist.gov/asd3 [2007, August 29]. National Institute of
  Standards and Technology, Gaithersburg, MD}.

\bibitem[{\citenamefont{Hunter et~al.}(1991)\citenamefont{Hunter, Krause,
  Berkeland, and Boshier}}]{stk1}
\bibinfo{author}{\bibfnamefont{L.~R.} \bibnamefont{Hunter}},
  \bibinfo{author}{\bibfnamefont{D.}~\bibnamefont{Krause}},
  \bibinfo{author}{\bibfnamefont{D.~J.} \bibnamefont{Berkeland}},
  \bibnamefont{and} \bibinfo{author}{\bibfnamefont{M.~G.}
  \bibnamefont{Boshier}}, \bibinfo{journal}{Phys. Rev. A}
  \textbf{\bibinfo{volume}{44}}, \bibinfo{pages}{6140} (\bibinfo{year}{1991}).

\bibitem[{\citenamefont{Moore}(1971)}]{NIST}
\bibinfo{author}{\bibfnamefont{C.~E.} \bibnamefont{Moore}},
  \emph{\bibinfo{title}{Atomic Energy Levels}}, vol.~\bibinfo{volume}{35} of
  \emph{\bibinfo{series}{Natl.\ Bur.\ Stand.\ Ref.\ Data Ser.}}
  (\bibinfo{publisher}{U.\ S.\ Govt.\ Print.\ Off.}, \bibinfo{address}{U.S.\
  GPO, Washington, D.C.}, \bibinfo{year}{1971}).

\bibitem[{\citenamefont{Johnson et~al.}(1983)\citenamefont{Johnson, Kolb, and
  Huang}}]{RPA}
\bibinfo{author}{\bibfnamefont{W.~R.} \bibnamefont{Johnson}},
  \bibinfo{author}{\bibfnamefont{D.}~\bibnamefont{Kolb}}, \bibnamefont{and}
  \bibinfo{author}{\bibfnamefont{K.~N.} \bibnamefont{Huang}},
  \bibinfo{journal}{At. Data Nucl. Data Tables} \textbf{\bibinfo{volume}{28}},
  \bibinfo{pages}{333} (\bibinfo{year}{1983}).

\bibitem[{\citenamefont{Volz and Schmoranzer}(1996)}]{Voltz}
\bibinfo{author}{\bibfnamefont{U.}~\bibnamefont{Volz}} \bibnamefont{and}
  \bibinfo{author}{\bibfnamefont{H.}~\bibnamefont{Schmoranzer}},
  \bibinfo{journal}{Phys.\ Scr. T} \textbf{\bibinfo{volume}{65}},
  \bibinfo{pages}{48} (\bibinfo{year}{1996}).

\bibitem[{\citenamefont{Derevianko et~al.}(1999)\citenamefont{Derevianko,
  Johnson, Safronova, and Babb}}]{Derevianko:pol99}
\bibinfo{author}{\bibfnamefont{A.}~\bibnamefont{Derevianko}},
  \bibinfo{author}{\bibfnamefont{W.~R.} \bibnamefont{Johnson}},
  \bibinfo{author}{\bibfnamefont{M.~S.} \bibnamefont{Safronova}},
  \bibnamefont{and} \bibinfo{author}{\bibfnamefont{J.~F.} \bibnamefont{Babb}},
  \bibinfo{journal}{Phys.\ Rev.\ Lett.} \textbf{\bibinfo{volume}{82}},
  \bibinfo{pages}{3589} (\bibinfo{year}{1999}).

\bibitem[{\citenamefont{Safronova et~al.}(2006)\citenamefont{Safronova, Arora,
  and Clark}}]{bin1}
\bibinfo{author}{\bibfnamefont{M.~S.} \bibnamefont{Safronova}},
  \bibinfo{author}{\bibfnamefont{B.}~\bibnamefont{Arora}}, \bibnamefont{and}
  \bibinfo{author}{\bibfnamefont{C.~W.} \bibnamefont{Clark}},
  \bibinfo{journal}{Phys. Rev. A} \textbf{\bibinfo{volume}{73}},
  \bibinfo{pages}{022505} (\bibinfo{year}{2006}).

\bibitem[{\citenamefont{Gunawardena et~al.}(2007)\citenamefont{Gunawardena,
  Elliott, Safronova, and Safronova}}]{Safronova:8s07}
\bibinfo{author}{\bibfnamefont{M.}~\bibnamefont{Gunawardena}},
  \bibinfo{author}{\bibfnamefont{D.~S.} \bibnamefont{Elliott}},
  \bibinfo{author}{\bibfnamefont{M.~S.} \bibnamefont{Safronova}},
  \bibnamefont{and}
  \bibinfo{author}{\bibfnamefont{U.}~\bibnamefont{Safronova}},
  \bibinfo{journal}{Phys.\ Rev.\ A} \textbf{\bibinfo{volume}{75}},
  \bibinfo{pages}{022507} (\bibinfo{year}{2007}).

\bibitem[{\citenamefont{Blundell et~al.}(1991)\citenamefont{Blundell, Johnson,
  and Sapirstein}}]{csao}
\bibinfo{author}{\bibfnamefont{S.~A.} \bibnamefont{Blundell}},
  \bibinfo{author}{\bibfnamefont{W.~R.} \bibnamefont{Johnson}},
  \bibnamefont{and}
  \bibinfo{author}{\bibfnamefont{J.}~\bibnamefont{Sapirstein}},
  \bibinfo{journal}{Phys.\ Rev.\ A} \textbf{\bibinfo{volume}{43}},
  \bibinfo{pages}{3407} (\bibinfo{year}{1991}).

\bibitem[{\citenamefont{Kreuter et~al.}(2005)\citenamefont{Kreuter, Becher,
  Lancaster, Mundt, Russo, H\"{a}ffner, Roos, H\"{a}nsel, Schmidt-Kaler, Blatt
  et~al.}}]{usca}
\bibinfo{author}{\bibfnamefont{A.}~\bibnamefont{Kreuter}},
  \bibinfo{author}{\bibfnamefont{C.}~\bibnamefont{Becher}},
  \bibinfo{author}{\bibfnamefont{G.}~\bibnamefont{Lancaster}},
  \bibinfo{author}{\bibfnamefont{A.~B.} \bibnamefont{Mundt}},
  \bibinfo{author}{\bibfnamefont{C.}~\bibnamefont{Russo}},
  \bibinfo{author}{\bibfnamefont{H.}~\bibnamefont{H\"{a}ffner}},
  \bibinfo{author}{\bibfnamefont{C.}~\bibnamefont{Roos}},
  \bibinfo{author}{\bibfnamefont{W.}~\bibnamefont{H\"{a}nsel}},
  \bibinfo{author}{\bibfnamefont{F.}~\bibnamefont{Schmidt-Kaler}},
  \bibinfo{author}{\bibfnamefont{R.}~\bibnamefont{Blatt}},
  \bibnamefont{et~al.}, \bibinfo{journal}{Phys.\ Rev.\ A}
  \textbf{\bibinfo{volume}{71}}, \bibinfo{pages}{032504}
  (\bibinfo{year}{2005}).

\bibitem[{\citenamefont{Johnson et~al.}(1988)\citenamefont{Johnson, Blundell,
  and Sapirstein}}]{johnson:Bspline88}
\bibinfo{author}{\bibfnamefont{W.~R.} \bibnamefont{Johnson}},
  \bibinfo{author}{\bibfnamefont{S.~A.}~\bibnamefont{Blundell}}, \bibnamefont{and}
  \bibinfo{author}{\bibfnamefont{J.}~\bibnamefont{Sapirstein}},
  \bibinfo{journal}{Phys. Rev. A} \textbf{\bibinfo{volume}{37}},
  \bibinfo{pages}{307} (\bibinfo{year}{1988}).

\bibitem[{\citenamefont{Sansonetti et~al.}(2005)\citenamefont{Sansonetti,
  Martin, and Young}}]{NIST-new}
\bibinfo{author}{\bibfnamefont{J.}~\bibnamefont{Sansonetti}},
  \bibinfo{author}{\bibfnamefont{W.}~\bibnamefont{Martin}}, \bibnamefont{and}
  \bibinfo{author}{\bibfnamefont{S.}~\bibnamefont{Young}},
  \emph{\bibinfo{title}{Handbook of basic atomic spectroscopic data}}
  (\bibinfo{year}{2005}), \bibinfo{note}{(version 1.1.2). [Online] Available:
  http://physics.nist.gov/Handbook [2007, August 29]. National Institute of
  Standards and Technology, Gaithersburg, MD}.

\bibitem[{\citenamefont{Stoicheff and Weinberger}(1979)}]{rben}
\bibinfo{author}{\bibfnamefont{B.}~\bibnamefont{Stoicheff}} \bibnamefont{and}
  \bibinfo{author}{\bibfnamefont{E.}~\bibnamefont{Weinberger}},
  \bibinfo{journal}{Can.\ J.\ Phys.} \textbf{\bibinfo{volume}{57}},
  \bibinfo{pages}{2143} (\bibinfo{year}{1979}).

\bibitem[{\citenamefont{Auzinsh et~al.}(2007)\citenamefont{Auzinsh, Bluss,
  Ferber, Gahbauer, Jarmola, Safronova, Safronova, and Tamanis}}]{7d}
\bibinfo{author}{\bibfnamefont{M.}~\bibnamefont{Auzinsh}},
  \bibinfo{author}{\bibfnamefont{K.}~\bibnamefont{Bluss}},
  \bibinfo{author}{\bibfnamefont{R.}~\bibnamefont{Ferber}},
  \bibinfo{author}{\bibfnamefont{F.}~\bibnamefont{Gahbauer}},
  \bibinfo{author}{\bibfnamefont{A.}~\bibnamefont{Jarmola}},
  \bibinfo{author}{\bibfnamefont{M.~S.} \bibnamefont{Safronova}},
  \bibinfo{author}{\bibfnamefont{U.~I.} \bibnamefont{Safronova}},
  \bibnamefont{and} \bibinfo{author}{\bibfnamefont{M.}~\bibnamefont{Tamanis}},
  \bibinfo{journal}{Phys.\ Rev.\ A} \textbf{\bibinfo{volume}{75}},
  \bibinfo{pages}{022502} (\bibinfo{year}{2007}).

\end{thebibliography}
\end{document}